\documentclass[sigconf]{acmart}

\usepackage{float}
\usepackage[normalem]{ulem}
\usepackage[disable]{todonotes}
\newcommand{\new}[1]{#1}
\newcommand{\old}[1]{}

\AtBeginDocument{%
  }

\copyrightyear{2026}
\acmYear{2026}
\setcopyright{cc}
\setcctype{by}
\acmConference[AHs 2026]{The Augmented Humans International Conference 2026}{March 16--19, 2026}{Okinawa, Japan}
\acmBooktitle{The Augmented Humans International Conference 2026 (AHs 2026), March 16--19, 2026, Okinawa, Japan}
\acmPrice{}
\acmDOI{10.1145/3795011.3795017}
\acmISBN{979-8-4007-2351-3/2026/03}

\begin{document}


\title{Robot-Wearable Conversation Hand-off for Navigation}

\author{Dániel Szabó}
\email{daniel.szabo@oulu.fi}
\orcid{0009-0003-7299-9116}
\affiliation{%
  \institution{University of Oulu}
  \city{Oulu}
  \country{Finland}
}

\author{Aku Visuri}
\email{aku.visuri@oulu.fi}
\orcid{0000-0001-7127-4031}
\affiliation{%
  \institution{University of Oulu}
  \city{Oulu}
  \country{Finland}
}

\author{Benjamin Tag}
\email{benjamin.tag@unsw.edu.au}
\orcid{0000-0002-7831-2632}
\affiliation{%
  \institution{The University of New South Wales}
  \city{Sydney}
  \country{Australia}
}

\author{Simo Hosio}
\email{simo.hosio@oulu.fi}
\orcid{0000-0002-9609-0965}
\affiliation{%
  \institution{University of Oulu}
  \city{Oulu}
  \country{Finland}
}
\renewcommand{\shortauthors}{Szabó et al.}

\begin{abstract}

Navigating large and complex indoor environments, such as universities, airports, and hospitals, can be cognitively demanding and requires attention and effort. While mobile applications provide convenient navigation support, they occupy the user's hands and visual attention, limiting natural interaction. In this paper, we explore \textit{conversation hand-off} as a method for multi-device indoor navigation, where a Conversational Agent (CA) transitions seamlessly from a stationary social robot to a wearable device. We evaluated robot-only, wearable-only, and robot-to-wearable hand-off in a university campus setting using a within-subjects design with N=24 participants. We find that conversation hand-off is experienced as engaging, even though no performance benefits were observed, and most preferred using the wearable-only system. Our findings suggest that the design of such re-embodied assistants should maintain a shared voice and state across embodiments.
We demonstrate how conversational hand-offs can bridge cognitive and physical transitions, enriching human interaction with embodied AI. 

\end{abstract}

\ccsdesc[500]{Human-centered computing~Natural language interfaces}
\ccsdesc[300]{Human-centered computing~Empirical studies in HCI}
\ccsdesc[300]{Human-centered computing~Empirical studies in ubiquitous and mobile computing}
\ccsdesc[100]{Human-centered computing~Ubiquitous and mobile devices}

\keywords{}


\maketitle

\begin{figure}[h!]
    \centering 
    \includegraphics[width=.48\textwidth]{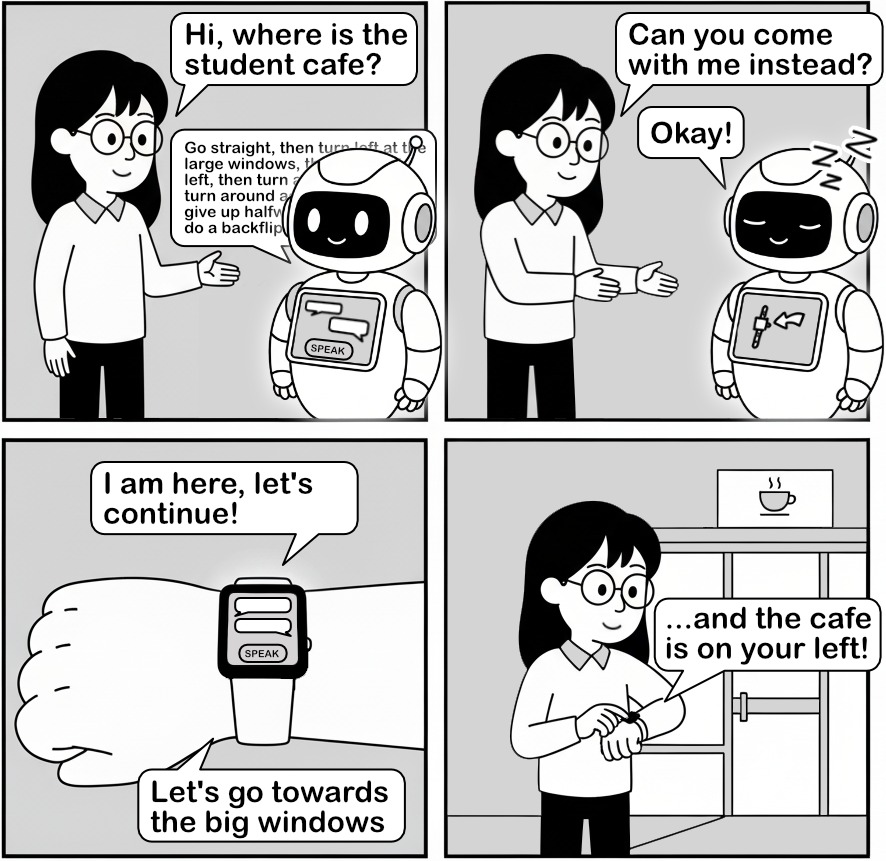}
    \caption{Usage of the conversation hand-off system presented as a storyboard, where the user looks for a cafe.}
    \Description{A four-panel cartoon storyboard illustrating a conversation hand-off for indoor navigation. In Panel 1 (Top Left), a user asks a white humanoid Pepper robot, "Hi, where is the student cafe?" The robot displays navigation instructions. In Panel 2 (Top Right), the robot's tablet shows "Okay!" In Panel 3 (Bottom Left), the user looks at a smartwatch displaying "I am here, let's continue!" while the robot in the background shows instructions. In Panel 4 (Bottom Right), the smartwatch displays the final navigation instruction: "...and the cafe is on your left!"}
    \label{fig:comic}
\end{figure}

\section{Introduction}

Indoor navigation aids are important for ensuring easy navigation in increasingly vast and complex indoor environments, such as universities, airports and hospitals \cite{Ivanov_2010, Plikynas_2020}. Given the challenges of navigating complex indoor spaces, technologies that provide flexible, context-aware guidance are particularly valuable, making Conversational Agents (CAs) a promising solution. Conversational Agents (CAs) offer several benefits in public service contexts, such as customer experience, service speed, and availability~\cite{Idris, Podpora_Gardecki_KawalaSterniuk_2019}. CAs have also shown potential in navigation assistance \cite{liu2024caven}.
CAs have an underexplored affordance in the context of navigation: their capability to work with various physical hardware, including mobile devices. 
In other words, they are not limited by space, hardware availability and can operate from multiple places and devices, even simultaneously. This flexibility allows CAs to be integrated with social robots; autonomous, embodied devices designed to interact with people in physical spaces. Embodiment within a social robot also increases a CA’s ability to engage people in interaction \cite{kiesler2008anthropomorphic}. This multi-modal presence potentially extends human perceptual and cognitive capacities across contexts and embodiments. By enabling assistance to ''follow`` the user beyond a single device, CAs act as an adaptive extension of the user’s spatial reasoning rather than a disconnected tool.

In 2001, \citeauthor{Butz_Baus_Krüger_Lohse_2001} described a hypothetical \textit{"Hybrid Indoor Navigation System"} \cite{Butz_Baus_Krüger_Lohse_2001} where a navigation assistant appears to the user in various ways depending on where they are in the building, as they proceed toward their destination. Digital navigation support has increasingly shifted toward multi-device and context-adaptive systems, reflecting long-standing HCI work on cognitive augmentation and distributed cognition \cite{schmidt2017augmenting, clinch2023augmented, hollan2000distributed}. Such systems illustrate how navigation support can become a form of situated augmentation, distributing information processing across devices, environments, and the user’s own cognitive strategies. In this framing, the environment itself becomes part of the user’s augmented cognitive system, enabling more fluid transitions between situated and mobile guidance.
This presents an opportunity to augment the user's navigation skills for indoor navigation more naturally, conveniently, and safely, addressed by prior studies that have shown the benefits of multi-embodiment systems for navigation \cite{caduff2008assessment, Butz_Baus_Krüger_Lohse_2001, pedestrian_nav_sys_13}. These multi-embodiment systems support the broader goal of augmenting human abilities by reducing cognitive load, enhancing environmental awareness, and offering continuous, adaptive scaffolding during spatial tasks.

In this paper, we explore \textit{conversation hand-off} as part of a navigation system, where the CA seamlessly switches between devices to provide a continued interaction. If successful, the hand-off can reduce mental effort in indoor navigation by allowing users to begin with a stationary device (social robot) and continue the task on a mobile device as they move toward their destination. This approach builds on existing models of wayfinding \cite{Dalton_Holscher_Montello_2019, He_McNamara_Bodenheimer_Klippel_2019, Marquardt_Greenberg_2012} and prior work demonstrating that mobile and wearable technologies can support cognitive enhancements in navigation and make them more accessible \cite{clinch2023augmented, schmidt2017augmenting}. By enabling a fluid, cross-device continuation of guidance, conversation hand-off operationalises a core principle of augmented humans: seamless integration between human cognition and distributed technological support. It allows navigation assistance to become pervasive, embodied, and sensitive to the user’s changing context, ultimately functioning as an adaptive augmentation of spatial cognition.
The scenario we built, depicted in \autoref{fig:comic}, explores such a conversation hand-off on a university campus. 

We investigate the hand-off concept as a mechanism for situated cognitive augmentation. Specifically, we examine how cross-device interaction influences perceptual engagement and supports efficient indoor navigation, while identifying how such systems can seamlessly extend human capabilities. We address this through three research questions:

\textbf{RQ1}: How does the hand-off of a human-AI conversation from a stationary embodiment to a wearable embodiment affect \textit{user experience}? 

\textbf{RQ2}: How does the hand-off of a human-AI conversation from a stationary embodiment to a wearable embodiment affect \textit{navigation performance}? 

\textbf{RQ3}: What \textit{design considerations} can we identify based on our experiment for an AI conversation hand-off between wearable and robot embodiments?

This paper's contribution is twofold. 
First, we present the implementation of the interaction paradigm itself. 
Our implementation uses a social robot and a smart watch, and is designed to assist humans in navigation tasks, including a seamless transition between the devices. This illustrates a concrete approach to augmenting humans’ navigational abilities through distributed and embodied AI assistance.
Second, we present our findings from a user study between three different embodiment configurations, including a set of design considerations grounded in reflection on the design process and qualitative analysis of the results from the user studies. These findings contribute to the broader discourse on augmented humans by showing how multi-device conversational agents can act as adaptive cognitive supports for real-world navigation tasks. 

\section{Background}

\subsection{Social Robots and Public AI Assistants}

The use of automated agents to provide information and assistance in public spaces is a growing area of research, encompassing social robots and public AI assistants. Social robots, which may be defined as \textit{machines that users may interpret as having social abilities}, have been studied extensively in fields such as education \cite{TanakaPepperApplication, Blair2023DevelopmentRobot, Mubin2020PepperImpressions, Matulik2020EdutainmentRobot, Onchi2016IntroducingEnvironment, Suddrey2018EnablingLaboratory, Chowdhury2020TheUniversity,sonderegger2022Enhancing}, hospitality \cite{Tuomi2021SpicingPepper, DeGauquier2018HumanoidShop, Gardecki2018TheApplication,lee2018technology,song2022Service}, and healthcare \cite{Blindheim2023PromotingStudy, Sato2020RehabilitationJapan, Uluer2020TowardsToday}. Within Human-Computer Interaction (HCI), their roles have been explored as library assistants, campus guides, and information providers \cite{Mubin2020PepperImpressions, Chowdhury2020TheUniversity, Blair2023DevelopmentRobot, Suddrey2018EnablingLaboratory}, which are examples of public AI assistants.

While the physical embodiment of a robot can enhance engagement, the design of the underlying conversational agent (CA) presents challenges that can hinder adoption. Evidence shows that while users often prefer CAs over traditional interfaces like websites for campus activities \cite{ca_for_smart_campus_20} or academic tasks \cite{virtual_assistant_for_students_20}, the implementation requires careful consideration. For example, a study by \citet{good_to_chat_20} on a social robot acting as a university host found little benefit from the deployed system. This research, however yielded three critical design recommendations for such public assistants: (i) the system's primary focus should remain on relevant information and tasks; (ii) open-domain social chat should be offered only on-demand rather than as a default; and (iii) topics should be appropriate for the spatial context, as users are less willing to share personal preferences in public. These principles were adopted in our work, as described in \autoref{sec:system}.

\subsection{Intelligent Navigation Systems}

Intelligent indoor navigation systems are an increasingly researched topic within the field of HRI. Recent work tackled a range of scenarios, humanoid mobile social robot behaviours for using elevators,
\cite{Gallo_2022}, wheeled non-humanoid robots \cite{Nanavati_2018} for guidance of visually impaired users and even unmanned aerial vehicles \cite{Bevins_2024}. \cite{Zahabi_2023} review shows a maturing field with an abundance of research and guidelines about the design and implementation of such a system; however, the field remains focused on visual and tactile interfaces. 

Meanwhile, AI assistants previously present in handsets have already made their way to consumer wearables such as the Apple Watch\footnote{\url{https://support.apple.com/en-gb/guide/watch/apd02f71f945/watchos}}, Galaxy Watch\footnote{\url{https://www.samsung.com/us/apps/bixby/}} and Fitbit\footnote{\url{https://assistant.google.com/platforms/wearables/}}.

Various kinds of wearable systems have been introduced to aid navigation \cite{swan_07, Chitra_Balamurugan_Sumathi_Mathan_Srilatha_Narmadha_2021, Dhivya_Premalatha_Monica_2019, Bouteraa_2023} and smart watches as navigation devices in particular have been received positively \cite{StepByWatch}. Although convenience and availability made mobile applications popular as navigation assistants \cite{hadwan2020towards, retscher2021development, secure_indoor_location_for_airports_18, pedestrian_nav_for_complex_public_facilities_16, pai2020smart}, handheld navigation divides the user's visual attention and can cause difficulty for users, particularly for those with physical or cognitive impairments \cite{Zahabi_2023}. Common limitation is their primary reliance on visual or touch-based interactions \cite{Perebner_19, SmartWatchNav_22, pedestrian_nav_sys_13}. \new{
Non-verbal audio interfaces were also considered, such as directionally annotated music \cite{Yamano_Hamajo_Takahashi_Higuchi_2012}, with the authors suggesting that a single-modal interface might be beneficial for mobile tasks.
}

Often overlooked, \textit{Landmark navigation} uses cues from the environment itself to guide the user's movement \cite{Sjolund_2018}. \citeauthor{pedestrian_nav_sys_13}'s mobile pedestrian navigation system \cite{pedestrian_nav_sys_13} compared a landmark-based and a street name-based version of a visual navigation application for a smart watch. The landmark-based approach was perceived to have significantly lower mental demand and effort required, using a NASA-RTLX scale (raw TLX). These findings highlight the viability and the benefits of landmark-based navigation.

\subsection{Re-Embodiment}

Agent chameleons and separating ``mind'' and ``body'' to allow agents to adapt to their physical forms, were discussed as early as 2003 by \citet{Duffy_OHare_Martin_Bradley_Schoen_Phelan}. 
Interest in multi-embodiment systems within HRI was renewed in 2019, as the work of \citet{Luria_Reig_Tan_Steinfeld_Forlizzi_Zimmerman_2019} lays down an important foundation for understanding the ways agents and embodiments can relate to each other in a multi-embodiment context. 
The authors carried out user studies on different variations such as re-embodiment, co-embodiment, one-for-one and one-for-all. 
\citeauthor{Luria_Reig_Tan_Steinfeld_Forlizzi_Zimmerman_2019} focus on re-embodiment, where a single agent inhabits multiple manifestations in succession, as exemplified by conversation hand-off. 
\citeauthor{uwa_22} studied a re-embodiment system consisting of a stationary and a wearable CA \cite{uwa_22}.  The results showed that users found the \textit{flexibility} of the ubiquitous system useful, but the speech input was implemented as push-to-talk, and some participants found this inconvenient compared to hotword detection (e.g. "Hey, Siri"), and some wished there was a keyboard-based input option for public use. 

Recently, a scoping review of multi-embodiment research by  \citeauthor{mind_body_identity_2024} identified that there is ``a significant gap in available guidelines for the design of safe, transferable, and scalable multi-embodied agent systems for real-world applications'' \cite{mind_body_identity_2024}.

Building on prior evidence of landmark-based navigation systems being positively perceived and effective in making complex spaces accessible, we extend this line of work by implementing a conversational agent-based navigation system for a public space.
Yet, existing research offers limited guidance on how such systems can be designed to remain flexible across multiple embodiments. To address this gap, our user study investigates participants' subjective experiences with a real-world, multi-embodiment conversational navigation system.

\section{Method}

To address our research questions, we designed, implemented, and evaluated a conversational navigation system that enables a hand-off between a stationary social robot and a wearable device. We conducted a within-subjects user study in a university campus setting to compare user experience and navigation performance across three conditions: using only the robot, only the wearable, and the robot-to-wearable hand-off. This section details the system's technical architecture, the embodiments' hardware, the design of the hand-off interaction, and the experimental protocol.

\subsection{System}
\label{sec:system}

In this section, we introduce each software module involved in the system, the hardware used for the study and the implementation of the conversation hand-off as motivated by our research questions.

\subsubsection{Server.}
\label{sec:server}

During our experiment, a local dedicated server performed all three stages of the speech-to-speech conversation pipeline and communication between the clients. The two clients that used this server are a humanoid social robot and a wearable device. The source code is publicly available
\footnote{\url{https://github.com/Crowd-Computing-Oulu/drs}}.

\paragraph{Speech Synthesis.}
We used Mimic 3 \cite{mimic3}, an open-source speech synthesiser model by Mycroft AI. Originally built for the embodied desktop assistant robot Mycroft Mark II \cite{mark2} and released as an independent module with an HTTP API and Docker support, Mimic 3 generates realistic speech audio from a text input. We chose Mimic 3 because it produces high-quality output and is meant to run as a service available to clients, making it ideal for sharing a voice between devices, a critical requirement for seamless conversation hand-off.

\paragraph{Speech Recognition.}
OpenAI's Whisper model \cite{robust_speech_recognition} offers accurate transcription of multi-language and noisy speech audio. \citet{whisper_cpp} implements the model that is fully open-source under MIT licence. Because Whisper.CPP, wrapped in another project Whisper.API \cite{whisper_api}, can be accessed as a HTTP API, similar to the speech synthesiser, further reducing computational load on the devices and enhancing response times. Like \citet{uwa_22}, speech input is implemented as push-to-talk.

\paragraph{Conversational Agent.}
Introduced in 2017 by \citet{RASA_2017}, Rasa made machine-learning-based dialogue management approachable to non-specialist software developers. 
Rasa supports calling arbitrary Python functions as part of the dialogue and is well-documented throughout, allowing us to implement a rigid logical process, navigation, with a flexible, natural conversational interface. As recommended by \citet{good_to_chat_20}, the system does not offer open-domain chat (as interactions are limited to greetings, navigation and the hand-off) and does not collect or store any personal information.

\subsubsection{Robot}
\label{sec:robot_client}
SoftBank Robotics' Pepper\footnote{\url{https://www.aldebaran.com/en/pepper}}\footnote{Our particular robot is version 1.8 with the NAOqi 2.9 Android operating system.} is a child-sized humanoid social robot that embodies the CA as a stationary, interactive and multi-modal device. It is widely used in education \cite{Tanaka_Isshiki_Takahashi_Uekusa_Sei_Hayashi_2015, Chowdhury2020TheUniversity, Blair2023DevelopmentRobot}, hospitality and retail \cite{Tuomi2021SpicingPepper, DeGauquier2018HumanoidShop, Gardecki2018TheApplication} and healthcare\cite{Blindheim2023PromotingStudy, Sato2020RehabilitationJapan, Uluer2020TowardsToday}.

Pepper's speak button is located on the chest display, which also shows the transcription of the conversation as chat bubbles. When Pepper speaks, it also performs a speech animation using its body. Otherwise, Pepper is completely stationary. The usage of the robot embodiment is shown in \autoref{fig:comic}, \autoref{fig:handoff_gui} and \autoref{fig:experiment_photos}.

Tapping the speak button on the robot's display starts an audio recording using the built-in microphones. On release, the recording is sent to the speech recogniser. When the transcription of the user's utterance is received, it is sent to the CA. As the response from the CA arrives, it is sent to the speech synthesiser. The CA response may also contain special instructions, such as the initiation of conversation hand-off. When the speech synthesiser returns the audio, it is played on the device's speakers.
This functionality is identical for the wearable device to ensure similar, intuitive controls for both. Throughout the implementation process, we aimed to minimise system response time (time between the end of the user's utterance and the beginning of the assistant's response) to make it viable for navigation tasks consistently across the input devices. The average response time is 2.89 seconds on average for the robot, 4.26 seconds for the watch, and 3.96 seconds for the conversation hand-off itself.

\subsubsection{Wearable Device}
\label{sec:wearable_client}

We used the Apple Watch Ultra 2 \cite{watch_ultra_2}, a smartwatch with a touchscreen, built-in speaker and microphone, and high-speed connectivity enabling fast audio transfer.

\begin{figure}[htbp!]
    \centering
    \includegraphics[width=0.48\textwidth]{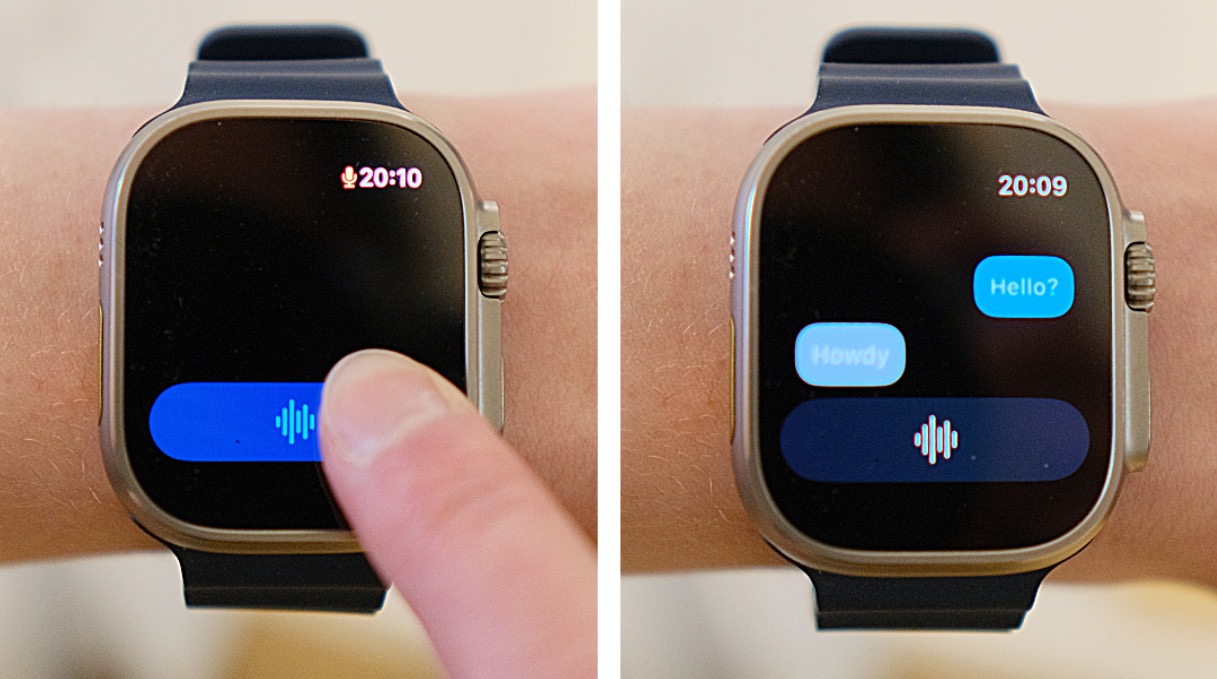}
    \caption{Photographs showing the user interface of the wearable navigation assistant embodiment. Holding the button while speaking (left), and when the processing is finished, the assistant is ready for another input (right).}
    \Description{Photographs showing the user interface of the wearable navigation assistant embodiment. Holding the button while speaking (left), and when the processing is finished, the assistant is ready for another input (right).}
    \label{fig:watch_photos}
    \vspace{-8pt} 
\end{figure}

We designed the system to be as simple as possible to make it easier to use while moving or being in a rush. As shown in \autoref{fig:watch_photos}, unlike the robot, it only shows the transcription of the last interaction to slightly reduce reliance on the textual transcripts during interactions and encourage prioritising the voice interactions for better spatial awareness. The system plays the assistant's synthesised speech without any additional animation.  

\subsection{Conversation Hand-off}
\label{sec:conv_handoff}

\begin{figure}[h!]
    \centering
    \includegraphics[width=.48\textwidth]{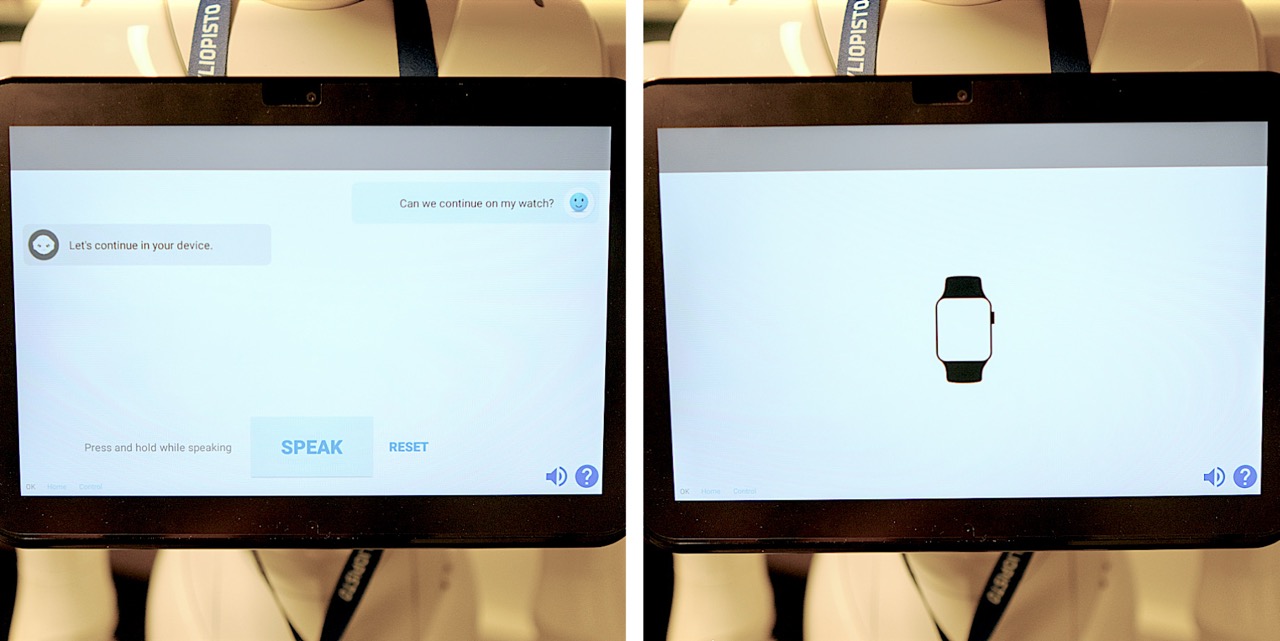}
    \caption{Conversation hand-off GUI on the Robot's chest tablet.}
    \Description{Two photographs of the robot's chest tablet. The first shows the user's input and the robot's response as chat bubbles on the screen, where the user asks the assistant to come along, and it confirms. The second shows the tablet displaying a watch icon.}
    \label{fig:handoff_gui}
    \vspace{-8pt} 
\end{figure}

Our two main design goals were to make the hand-off intuitive and pleasant. 
The trigger of the hand-off procedure was a voice command, such as "Can we continue on my watch?", "Can you come with me?" or "Hop over to my watch".
Alternative designs considered were proximity-based or implicit (simply talking to the other device), but we decided that the ritual of the explicit hand-off process may help the participants develop an accurate mental model of the system.
To maximise system transparency, both devices confirm their states, and the CA clearly prompts the user on what to do next. 
As shown in \autoref{fig:comic} and \autoref{fig:handoff_gui}, the user asks for the hand-off, which is then confirmed by the CA on the robot before performing it. Then, the robot displays a watch icon, and the CA verbally checks in on the watch, saying, "Hi, I'm here. Let's continue.". 

\subsection{Experimental Design}

We conducted a within-subject experiment in a controlled environment to assess the developed system and participants' experience and performance.

Each experiment lasted one hour for each participant.  The exact procedure is detailed in \autoref{exp_procedure}.  We conducted the experiments during times of minimal traffic (late afternoons and weekends) in the experimental areas to minimise disturbances between participants. The experiment area was a large open space with multiple corridors and smaller obstacles, to mimic a real navigation task in a scaled-down format. 


\subsubsection{Conditions.}

A balanced Latin square design ensured equal participant exposure to different conditions and assignments of conditions to routes while mitigating order effects. 

\textbf{Robot-Only Navigation (Robot)}: Participants exclusively relied on navigation instructions provided by the Pepper robot. The robot remained stationary during navigation.

\textbf{Wearable-Only Navigation (Wearable)}: Participants solely utilised the Apple Watch for navigation guidance.

\textbf{Robot-to-Wearable Hand-off (Hand-off)}: Participants commenced navigation with the Pepper robot and transitioned to the Apple Watch during the course of navigation.

\subsubsection{Setting.}

The study was conducted within the university campus environment. In three conditions \textit{Robot}, \textit{Wearable}, and \textit{Hand-off}, participants navigated three predetermined routes. Routes were assigned to conditions in all possible combinations to minimise the potential effects of the routes on the conditions. Each of the routes comprised four landmark checkpoints, the last one being the destination. The landmarks are brightly coloured $100mm*100mm$ geometric shapes (red, green or blue coloured square, disk or triangle shape) stuck on the wall at eye level ($160cm$), as seen in \autoref{fig:experiment_photos}. The routes were designed to be comparable in their difficulty.
The routes and the overall experiment layout are shown in Appendix G. The coloured geometric shapes were used to reduce ambiguity in the navigation tasks, to reduce ability differences in language or vision and because among the participants were students who may have been familiar with the building in which the experiment took place. To avoid participants navigating from their memory of shapes, we strategically placed duplicate shapes as well.

\subsection{Participants}

For this study, we recruited $N = 24$ participants, mostly students and university staff, offline or via a university-wide human subject pool managed by the university's research services department. All participants were required to have normal or corrected-to-normal vision, as well as unimpaired speech, hearing or mobility.

\subsection{Procedure}
\label{exp_procedure}

Each experiment consisted of three main parts and was designed to last approximately 60 minutes. Our participants were compensated with university-branded merchandise valued at €20.

\begin{figure}[htbp!]
    \centering
    \includegraphics[width=0.48\textwidth]{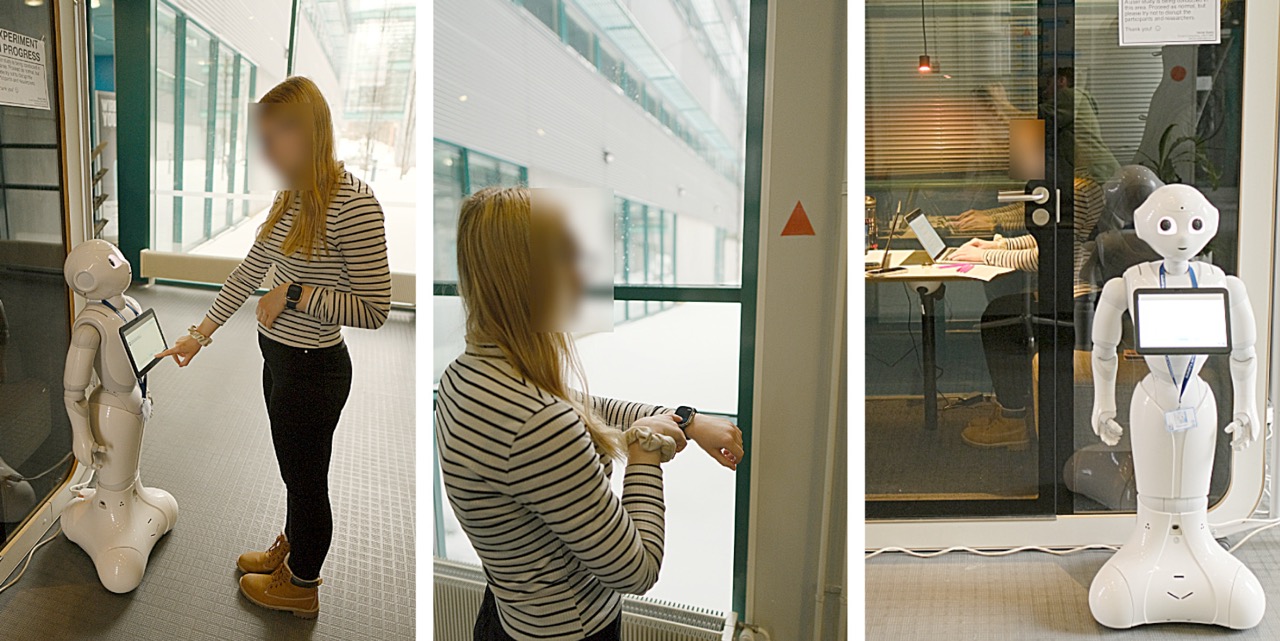}
    \caption{The participant uses the robot situated in front of the booth to begin navigation (left) and uses the watch as she makes her way through the landmarks (middle) and fills out questionnaires at the start point before, between and after the three conditions right).}
    \Description{The participant uses the robot situated in front of the booth to begin navigation (left) and uses the watch as she makes her way through the landmarks (middle) and fills out questionnaires at the start point before, between and after the three conditions right).}
    \label{fig:experiment_photos}
\end{figure}

The first part began by familiarising the participant with the study and the procedure. Then, the participant filled out a demographic questionnaire (see \hyperref[sec:appendix_a]{Appendix A}), including questions regarding subject familiarity. 

The second part was the experiment itself. This step was repeated three times for each participant, with conditions and routes according to our balanced Latin square design. Participants were familiarised with the tasks and devices using an instruction sheet and by trying the devices themselves. They navigated with the devices' help (illustrated by \autoref{fig:experiment_photos}). After each task, the participants filled out the per-condition questionnaire (see \hyperref[sec:appendix_b]{Appendix B}) with questions regarding workload, trust in the system and general feedback on the experience. 

Finally, the participant filled out a final post-experiment questionnaire (see \hyperref[sec:appendix_c]{Appendix C}) with questions regarding the participant's preference between navigation methods and specific questions regarding the conversation hand-off conditions.

\subsection{Measurements}

We collected eight types of data during the experiments. See \hyperref[sec:appendix_b]{Appendix B} for the per-condition questionnaire and \hyperref[sec:appendix_c]{Appendix C} for the post-experiment questionnaire.
To answer RQ1, we used the NASA-RTLX~\cite{nasatlx} and the Trust in System \cite{Jian_Bisantz_Drury_2000} scales and collected open feedback in all questionnaires. We also asked for users' preferences among the three embodiment configurations and justifications for their choices.
To help answer RQ2, we measured accurate task completion times, error rates, interaction counts and collected researcher notes that helped summarise the types and causes of any navigation mishaps or unexpected interactions.

\begin{itemize}
    \item \textbf{Workload} The six-item NASA Task Load Index (NASA-RTLX) questionnaire \cite{nasatlx} is filled out after each condition, with each item being a 7-point Likert scale from Strongly Disagree to Strongly Agree. NASA-RTLX helps us understand whether the hand-off creates significant additional workload (cognitive or otherwise) that could impact its user experience, and ultimately its adoption.
    \item \textbf{Trust in system} The two-part, twelve-item questionnaire \cite{Jian_Bisantz_Drury_2000} is filled out after each condition, with each item being a 7-point Likert scale from Strongly Disagree to Strongly Agree. We collect this data because trust has been identified as a key component of technology adoption \cite{Jian_Bisantz_Drury_2000}.
    \item \textbf{Task Completion Time} The time it takes for participants to reach their destination from the starting point, manually measured in seconds by the researcher.
    \item \textbf{Error Rate} The researcher records, based on observation, whether the participant significantly deviated (enough to have to return to a previous landmark) from the intended path at least once.
    \item \textbf{Interactions} The researcher records, based on observation, the number of any user input made.
    \item \textbf{User Preference} The participants fill out questions regarding their preferences as part of the post-experiment questionnaire. 
    \item \textbf{Open Feedback} Participants provide open-ended feedback regarding their experience with the system, after each condition and at the end of the experiment (see \autoref{sec:appendix_b} and \hyperref[sec:appendix_c]{C} ). Additionally, they are asked to describe the conversation hand-off experience.
    \item \textbf{Researcher Notes} The researcher takes note of unexpected user behaviours.
\end{itemize}

\subsection{Data Analysis Methodology}

All quantitative data were processed in R version 4.4.1. and we followed standard statistical methods in HCI, first testing for data normality using a Shapiro–Wilk test and then selecting appropriate subsequent methods. Because our data were all either ordinal or non-normal, we used Friedman tests to make overall comparisons and Wilcoxon post-hoc pairwise comparisons where the omnibus test detected differences.
We provide additional details on the particular analyses in the Results Section (\autoref{sec:results}).

For our qualitative data analysis, we followed the principles of descriptive qualitative analysis. 
The lead author formulated an initial set of codes, then two authors iteratively improved the codes by merging and splitting them. 
Thematic Analysis \cite{clarke2014thematic} was performed separately for the per-condition questionnaire responses and for the post-experiment questionnaire. The final code set for the per-condition data is shown in \hyperref[sec:appendix_e]{Appendix E} and for the post-experiment data in \hyperref[sec:appendix_f]{Appendix F}.

\section{Results}
\label{sec:results}

\subsection{Quantitative Findings}

\subsubsection{Demographics}

Among our 24 participants, 58.3\% (N=14) identified as male and 41.7\% (N=10) as female. 
Ages ranged narrowly (\textit{M} = 26.2, \textit{SD} = 4.3). Participants from Asian countries comprised 58.3\% (N = 14), Europeans 16.7\% (N = 4), and other regions 25\% (N = 6).
Familiarity (5-point scale: 1 = not familiar, 5 = very familiar) was \textit{M} = 2.00 (\textit{SD} = 1.10) for social robots, 3.33 (\textit{SD} = 1.27) for AI assistants, and 3.29 (\textit{SD} = 1.20) for indoor navigation software. 
Reliance on any help for indoor navigation scored 3.79 (\textit{SD} = 0.98).
Use frequency (5-point scale: ``less than once a year'' to ``at least once a day'') showed the following patterns. 
For social robots, one participant reported weekly use, one monthly, four yearly, and 18 less than yearly. 
For AI assistants, six reported daily use, six monthly, three yearly, and two less than yearly. 
For indoor navigation, three reported daily use, eight weekly, eight monthly, three yearly, and two less than yearly.

\subsubsection{User Experience.}


\paragraph{Preference.}

Participants shared their preferences among the three systems they tried -- 70.8\% (N = 17) chose the wearable-only system, 29.2\% (N = 7) preferred the hand-off, and none preferred using only the robot. 

\paragraph{NASA-RTLX.}\label{sec:workload}

\begin{figure}[htbp!]
    \centering
    \includegraphics[width=.48\textwidth]{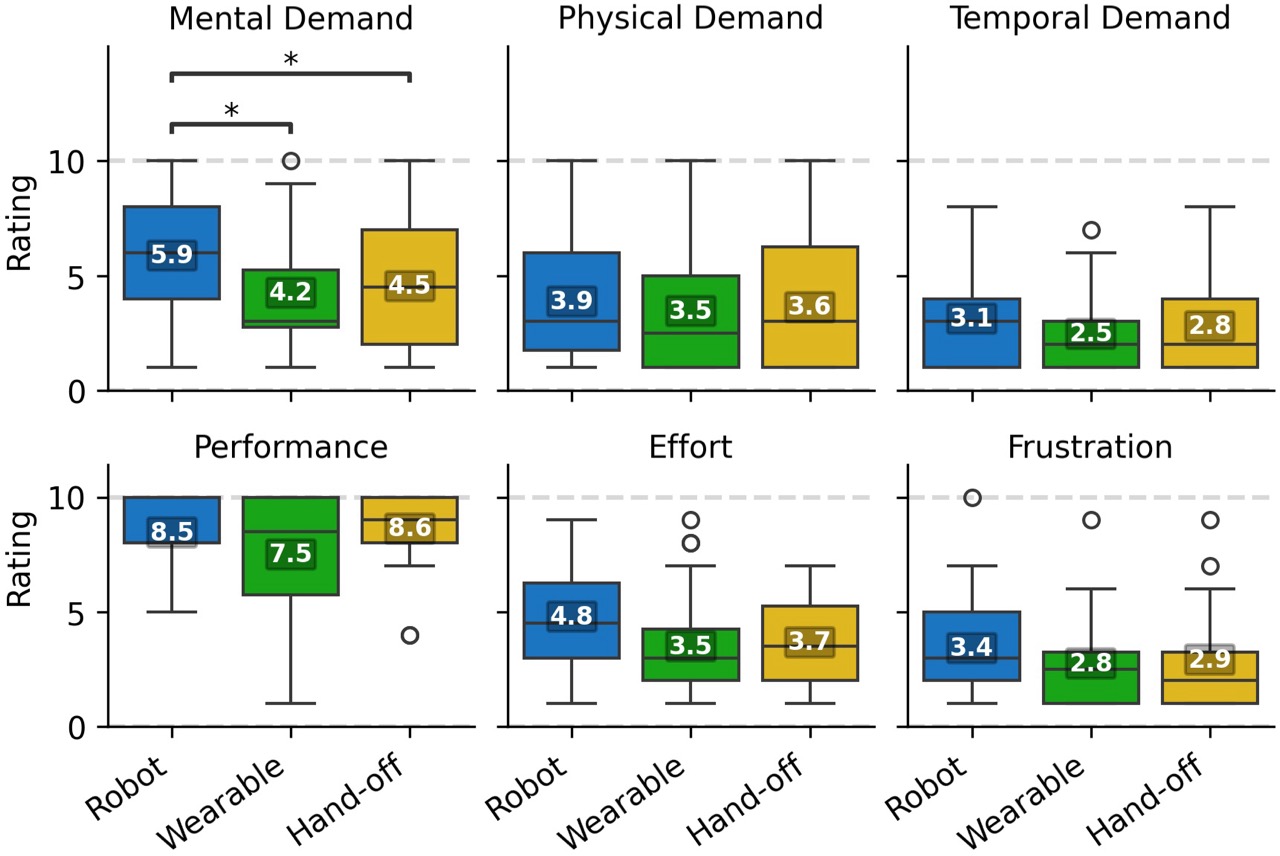}
    \caption{Box plot comparing NASA-RTLX scores for conditions.}
    \Description{Six box plots comparing NASA-RTLX scores across Robot, Wearable, and Hand-off conditions for Mental Demand, Physical Demand, Temporal Demand, Performance, Effort, and Frustration. For Mental Demand, Robot (median ~5.9) is higher than Hand-off (~4.5) and Wearable (~4.2). For Physical Demand and Temporal Demand, all conditions show similar low medians (~2.5-3.9). For Performance, all conditions show similar high medians (~7.5-8.5). For Effort, Robot (median ~4.8) is higher than Hand-off (~3.5) and Wearable (~3.7). For Frustration, all conditions show similar low medians (~2.8-3.4).}
    \label{fig:nasa_boxplot}
\end{figure}

Participants completed the NASA Task Load Index (NASA-RTLX)\cite{nasatlx} questionnaire after each navigation task to help us understand how the workload differs (\autoref{fig:nasa_boxplot}). 

We conducted a series of Friedman tests to compare the perceived workload across the Robot, Wearable, and Hand-off conditions for each of the six NASA-RTLX subscales.
The omnibus test detected a significant effect of condition on \textbf{Mental Demand}, $\chi^2(2) = 8.95$, $p = .011$. Post-hoc pairwise comparisons using the Wilcoxon signed-rank test with Bonferroni correction revealed that the Mental Demand in the Robot condition ($M=5.88$, $SD=2.35$) was significantly higher than in both the Wearable condition ($M=4.25$, $SD=2.72$), $p = .018$, and the Hand-off condition ($M=4.54$, $SD=2.83$), $p = .034$.
A significant difference was also found for perceived \textbf{Effort}, $\chi^2(2) = 13.85$, $p < .001$. However, the post-hoc comparisons did not detect any significant pairwise differences (all $p > .05$).
There were no significant differences detected for the remaining subscales: \textbf{Physical Demand} ($\chi^2(2) = 1.08$, $p = .583$), \textbf{Temporal Demand} ($\chi^2(2) = 5.64$, $p = .060$), \textbf{Performance} ($\chi^2(2) = 3.11$, $p = .211$), and \textbf{Frustration} ($\chi^2(2) = 4.03$, $p = .133$).

\paragraph{Measuring Trust.}

\begin{figure*}[b!]
    \vspace{0px}
    \centering
    \includegraphics[width=.95\textwidth]{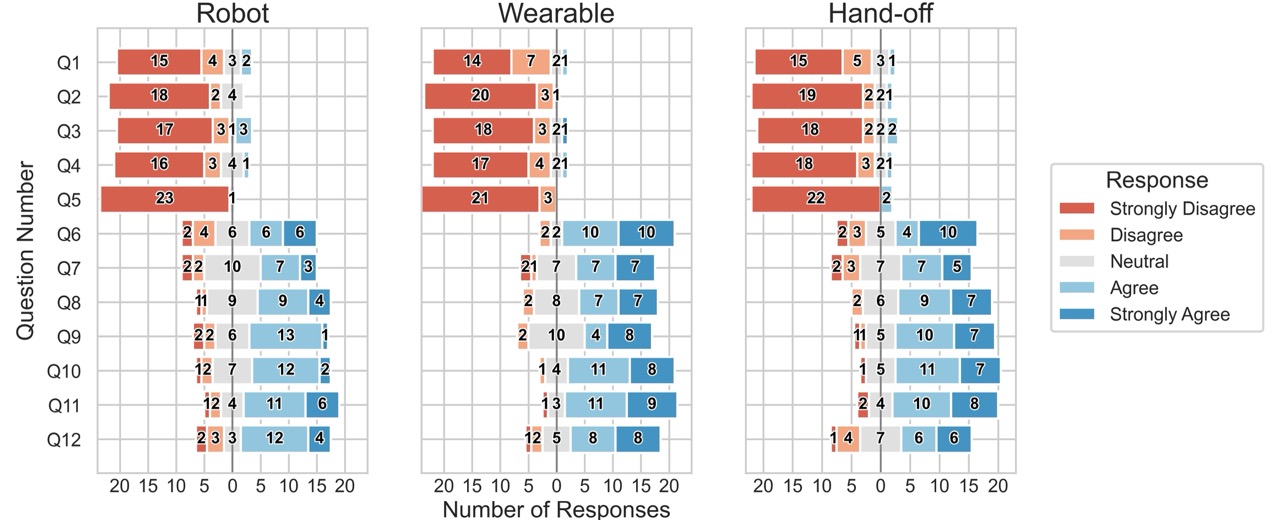}
    \caption{Comparing results from Trust Between People and Automation Scale \cite{Jian_Bisantz_Drury_2000} questions for each of the three conditions. See the questions in \autoref{sec:appendix_b} (items B8-B19).}
    \Description{Three vertical histograms displaying the distribution of responses for Trust Between People and Automation Scale \cite{Jian_Bisantz_Drury_2000} questions (Q1-Q12) across Robot, Wearable, and Hand-off conditions. Each chart uses a 5-point Likert scale from "Strongly Disagree" to "Strongly Agree," with counts for each response type, with at least one response for each question. Q1 to Q5 are overwhelmingly leaning to "Strongly disagree", while Q6 to Q12 show many neutral and slightly agree answers. The Wearable and Hand-off conditions appear to have more "Strongly Agree" responses from Q7 to Q12 than the Robot condition.}
    \label{fig:trust_boxplot}
\end{figure*}

The Trust Between People and Automation Scale \cite{Jian_Bisantz_Drury_2000} was filled in after each navigation task to gauge participants' perception of the system beyond pure utility.
Overall Friedman tests did not detect differences between conditions, and the marginal variance between conditions can be seen in~\autoref{fig:trust_boxplot}.

\subsubsection{Navigation Performance.}

We present quantitative data about the system and navigation performance. 


\paragraph{Task Completion Time.}\label{sec:task-completion-time}

We find no statistical significance (\textit{p} = .74) in task completion time, and the average task completion time for the experiment was 146.4 seconds. 
Comparing experiments chronologically, there was no significant learning effect observable (\textit{p} =  .78) and the difference between routes was also not significant (\textit{p} =  .54), as confirmed by a Kruskal-Wallis test.

\begin{table}[htbp]
\centering
\small
\caption{A Comparison of Performance Metrics Across Different Experiments, Conditions, and Routes.}
\label{tab:performance_metrics}
\begin{tabular}{@{}lrrr@{}}

& \multicolumn{3}{c}{\textbf{Conditions}} \\ 
\cmidrule(l){2-4}
\textbf{Metric} & \multicolumn{1}{c}{\textbf{Robot}} & \multicolumn{1}{c}{\textbf{Wearable}} & \multicolumn{1}{c}{\textbf{Hand-off}} \\ \midrule
Mean Task Time (sec) & 140.8 & 157.6 & 192.5 \\
Task Time SD (sec) & 57.9 & 79.6 & 88.2 \\
Error Rate (\%) & 62.5\% & 50.0\% & 50.0\% \\
Mean Interactions & 2.9 & 3.7 & 4.2 \\
Interaction SD & 1.9 & 1.2 & 2.1 \\
& \multicolumn{3}{c}{\textbf{Routes}} \\ 
\cmidrule(l){2-4}
\textbf{Metric} & \multicolumn{1}{c}{\textbf{Blue Square}} & \multicolumn{1}{c}{\textbf{Green Circle}} & \multicolumn{1}{c}{\textbf{Red Triangle}} \\ \midrule
Mean Task Time (sec) & 173.9 & 145.8 & 171.2 \\
Task Time SD (sec) & 79.8 & 84.1 & 70.5 \\
Error Rate (\%) & 12.5\% & 75.0\% & 75.0\% \\
Mean Interactions & 3.3 & 3.4 & 4.0 \\
Interaction SD & 2.1 & 1.6 & 1.7 \\
\end{tabular}
\end{table}

\paragraph{Error rate.}

There is no significant difference between the error rates while using different systems, and there is no discernible learning effect either. There is some difference between error rates on different routes, though: Route 1 to the Blue Square has a somewhat lower error rate, which may be explained by this being the only route where the destination is visible from further away and is part of other routes.

\paragraph{Interactions.}

In the case of the Hand-off, the input requesting transfer to the wearable was also included, which is reflected in the slightly higher count (\textit{M} = 4.2, \textit{SD} = 2.1) compared to the other two conditions (\textit{M} = 2.9, \textit{SD} = 1.9 and \textit{M} = 3.7, \textit{SD} = 1.2 respectively). These differences were, however, not statistically significant, as confirmed with a Kruskal-Wallis test (\textit{p} =  .24).

\subsection{Qualitative Findings}
\label{ssec:condition_feedback}

We present the results of qualitative analysis of participant feedback data, collected after each condition (see \autoref{sec:appendix_b}) and at the end of the experiment (see \autoref{sec:appendix_c}).
We include the used code set in \autoref{sec:appendix_e}. 
For each condition, we detail the positive feedback first, then the negative feedback, and finally, all other remarks. In this section, we refer to participants by their numbers from P1 to P24.

\subsubsection{Robot.}

Participants (P1, P11, P12) highlighted that the robot showing the verbal output as text on the screen was helpful, and almost all participants praised it for its clear and detailed directions. P23 particularly enjoyed the voice-based interaction as there is \textit{``no need to indicate my start point or target point in a huge messy map''}.
P5 highlighted that being able to request all the directions at once instead of step-by-step is helpful for them. P1 also mentioned that the instructions staying on the screen were helpful when they returned to the robot to re-read the output. A few participants asked for step-by-step directions from the robot and found it difficult to use (P2, P10), and others reported having to break down the navigation task into smaller tasks and return to the robot for the next section each time (P3, P4, P5).
Two participants also complained about speech recognition issues (P7, P10, P18) and that using the robot is too time-consuming (P8). To help with this, P9 also suggested showing the shapes visually. Further, P6 highlighted that Pepper is \textit{``not easy to carry''} as a downside. 

\subsubsection{Wearable.}

Almost all participants mentioned the benefit of being able to ask for directions at any point and generally liked the embodiment's portability. P2 and P10 mention that the wearable embodiment is particularly easy to use. P8 highlighted that it also provided a sense of privacy. 
P3 and P5 thought the assistant's directions were not precise and detailed enough, which made navigation more difficult in some cases. P4 thought the assistant was difficult to use because it used different commands than the Hand-off condition, even though all the devices embodied the same assistant, and this was mentioned to the participants. P9 found that they had to read too much to get to the destination. P2 recommended showing the entire conversation history just like on the robot's screen and using automatic positioning instead of telling the assistant where they are. Finally, P12 suggested hotword detection instead of push-to-talk. P20 suggested that the system would have benefited from visual directions, too.
P13 felt that with this system, successful navigation hinges on the user's ability, rather than the system's instructions: \textit{``If we follow the commands properly, it won't be difficult to use this system for navigation.''}. 

\subsubsection{Conversation Hand-off.}

Much of the feedback for participants is aimed at the assistant itself or the devices used. P4 highlighted that they thought it was very convenient to have access to the watch on the go if you forget the initial directions. Many (P1, P2, P3, P7, P12) thought that the directions were particularly good this time. The conversation hand-off itself received mostly positive remarks (P2, P9, P11, P12). P2 mentioned that the hand-off was smooth, and P9 was simply glad that the hand-off was an option. P12 said that \textit{``The transfer command is really good because I do not need to go back to the robot. The instructions from the watch were clear and not fast.''}. P11 found a specific aspect of the hand-off fun: \textit{``My favourite part was when the assistant said hello when moving from the robot to the wearable device.''}. P6 reported that after the hand-off process, it was \textit{``annoying having to repeat the question after moving the consciousness to the watch''}.

P3 and P12 reported issues with the assistant's memory of their location and destination not persisting across the devices. P9, P10 and P18 noted that the assistant talked too fast, and they would have preferred a slower, more \textit{``convenient pace''} (P9). P17 was glad that the robot's speech was slow (both devices used the same speech synthesiser, thus the speed was identical). Many (P1, P5, P7, P9, P15, P19) reported speech recognition issues, and a few praised the voice recognition's reliability (P10, P20). P3 and P12 found the directions lacking in clarity. P1 reported that \textit{``The system is quite slow at times.''}.
P8 mentioned that the navigation had a somewhat difficult start, but then it went fine afterwards: \textit{``once it understands where you are and where you have to go, then it works fine''}. P1 also reported improvement in their ability to use the system over time. P9 recommended hotword detection once again. P21 and P13 appreciated the ability to try again if they made mistakes or misunderstood the instructions. P22 wished the system were more proactive in offering flexible options: \textit{``The system could ask back if I want the whole route.''}. Two participants, P13 and P16, wished the system worked with languages other than English.

\subsubsection{Post-experiment Questionnaire} \label{ssec:post_experiment}

In this section, we present the analysis of the post-experiment questionnaire (see Appendix C).

\paragraph{Preference.}

Seven participants (P4, P5, P12, P13, P14, P15 and P22) preferred the conversation hand-off, and all others preferred the watch. Almost all who preferred the watch motivated their decision with the ability to be on the move while navigating or seeing no reason to use two devices for the same task, and if they can choose between the wearable and the robot, they pick the wearable (P1, P7). P8 saw the unique benefits of the wearable embodiment. They felt that the watch provided a sense of ownership and confidence compared to the robot. P5 suggested that the watch would be even more useful if it used some positioning technology to determine the user's location.

Those who preferred the Hand-off had various motivations. P4 identified the benefits of the two kinds of embodiments: \textit{``I like the human element of talking to the robot, but then not being limited by its lack of movement.''}. Further, P4 thought the conversation hand-off was \textit{``futuristic''} and that it made the task \textit{``more pleasant''} and felt \textit{``natural''}. On the other hand, P4 still thought the hand-off could be made smoother in future iterations. P12 described that they felt anxious using just the watch and felt better initiating the conversation on the robot and then transferring it to the wearable device. Similarly, P22 described a feeling of being \textit{``more connected to the AI''}.

P4 recommended making the transition between devices more seamless. Many recommended improving performance: either speech recognition (P7, P8, P10) or the AI assistant's flexibility (P1) or better consistency of instructions (P2). Other recommended features include a visual map (P10, P21), landmarks as a list (P3), chat history on the wearable device (P2), and hotword detection (P12). Few even recognised the inherent weaknesses of the embodiment types. P5 found that they did not like that they had to bend down to interact with the robot (which is child-sized) and that the robot did not move, and P11 and P24 thought the robot would be more enjoyable to use, if its functionality matched the wearable: \textit{``Talking with the watch is easy, but not very inspiring, unlike talking with the robot''} (P11). P22 suggested that \textit{``The robot could ask more questions or suggestions like 'Do you want to transfer to the watch', or 'Do you want the whole route or step by step instructions'.''}

\paragraph{Conversation Hand-off.}
The hand-off condition was seen to be just as good as the robot, but it had extra steps, making the robot unnecessary (P1, P2, P9, P10, P24). P11 thought the robot assistant created a \textit{``friendly atmosphere''} and that it \textit{``makes you feel more secure''}, but since it is static, it may become somewhat annoying to use. P9 and P12 both outlined that their willingness to initiate the interaction with the robot is dependent on the scenario. P9 describes the wearable being more useful if they are in a hurry at the moment, and the hand-off taking the extra time they cannot afford but wrote that if they are not in a hurry and are navigating \textit{``leisurely''}, they would like to interact with the robot, highlighting body gestures as an affordance of this type of embodiment. P12 explained that the robot is useful if they are unfamiliar with the location but would otherwise resort to the wearable assistant.

Many (P4, P10, P11) found the hand-off \textit{fun}: \textit{``That was my favourite experience. It was like the assistant was teleporting from one place to another''} (P11). Meanwhile, others (P9, P22) described it as \textit{convenient}. P6 thought the idea of having an AI assistant on a wearable device was mostly interesting and easy. P7, on the other hand, enjoyed the real-time nature of the interaction: \textit{``I personally enjoyed this technique since we can have live data. Not like taking instructions from the robot - we can identify the locations with live data which will lead to having more accuracy''}. P12 wrote that the hand-off made the navigation easier and that they could imagine the robot talking to them through the wearable device. Similarly, P9 also mentioned that they would like to download the assistant as an application to their phone. P4 thought this system even felt more natural than the other two. P5 also mentioned that even though the assistant would sometimes misunderstand them, they are fine with such mistakes. 
Participants made recommendations to improve the hand-off procedure. P17 wanted the hand-off to be faster and easier. P22 explained that the robot should proactively ask whether to make the hand-off, and wished that the hand-off was \textit{``more visually attractive''}.

\section{Discussion}

We conducted a user experiment investigating the transfer of a navigation CA from a stationary robot to a smartwatch via a conversation hand-off. 
First, RQ1 tackled the user experience of the hand-off. While the wearable-only condition was preferred by most for its convenience, the hand-off was seen as engaging and fun, a quality often leveraged in public AI assistants \cite{Chowdhury2020TheUniversity, Tuomi2021SpicingPepper}. The robot-only condition had a higher mental demand compared to both the wearable and hand-off conditions, which aligns with previous findings \cite{pedestrian_nav_sys_13}. User trust in the system was not negatively affected by the hand-off.
Then, RQ2 investigated the effect on task performance metrics. 
We found no statistically significant differences in task completion time, error rates, or interaction counts across the three conditions, indicating that the conversation hand-off is a viable interaction method that does not introduce a performance penalty compared to single-device solutions.
Next, we discuss experiment observations and present design considerations emerging from the study, answering RQ3.

\subsection{Roles of Social Robots and Wearable Devices in Navigation}

Based on our observations during experiments, we identified user strategies aimed at improving interaction efficiency. Participants consistently sought to minimise interaction frequency, using complex single-utterance commands (e.g., combining a greeting and destination) and requesting complete route information upfront. 
The latter behaviour was unexpectedly carried over from the robot to the wearable device. 
In the Robot condition, this preference for upfront information established the robot as a static "hub" for cognitive offloading; users would display the full route, memorise it in segments, and physically return to the robot to read the next steps. 

We observed a potential cognitive bias where participants, rather than recognising their own learning, anthropomorphised the system by attributing their own improved inputs to the AI "figuring out" their intentions.
Finally, the timing of the conversation hand-off illuminates the distinct perceived roles of the devices. 
Participants consistently used the robot to ground the task—initiating the transfer to the wearable device at the precise moment the robot confirmed its understanding by beginning to provide directions—before switching to the mobile device for the navigation itself. This form of first planning (AI) and then executing (human) has been shown to reduce cognitive load and increase trust in collaborative tasks with AI agents \cite{he2025plan}, although one should be wary of plausibly seeming but unrealistic plans. One of the older HCI theories explaining similar phenomena is the concept of \textit{distributed cognition} \cite{hollan2000distributed}, which has offered guidelines in the design of human-computer technologies. One of the earliest works by Butz et al. \cite{Butz_Baus_Krüger_Lohse_2001} already recognised the improvements distributed and hybrid systems can offer to human cognitive tasks.

The behaviours of our participants suggest they were actively distributing cognitive responsibilities across themselves, the robot and/or the wearable. This indicates that multi-embodied conversational systems can augment human navigation by structuring planning and execution across devices in a way that shapes and supports users’ own decision-making processes.

These findings highlight how each configuration may have distinct real-world beneficiaries. The \textit{Robot-only} setting serves as a high-visibility social anchor in public spaces like transit hubs or hospitals, benefiting first-time visitors who require an obvious starting point for orientation. The \textit{Wearable-only} setting caters to power users and daily commuters who value efficiency and minimal friction in familiar environments. Finally, the \textit{Hand-off} condition addresses the needs of users in complex, multi-stage journeys—such as tourists in a large museum or patients in a sprawling medical campus—who benefit from the initial high-bandwidth social engagement of a robot to ground the task, followed by the continuous, hands-free support of a wearable during the execution phase.

\subsection{\new{Conversation Hand-off as a Proxemic Interaction}}

The observed structuring of planning and execution across devices may be explained by the concept of \textit{Proxemics}. Proxemics concern the role of personal space in social interactions \cite{hall1966hidden}, and Proxemic Interactions in HCI \cite{Marquardt_Greenberg_2012} borrow principles Proxemics to design systems that are socially more ergonomic. \citeauthor{hall1966hidden}'s Proxemic Zones describe how even just one simple factor, distance, can affect how the user relates to another person or system, highlighting how our designed system can be viewed through the lens of Proxemics. As the user first enters the building, they discover the social robot (the \textit{public zone}), then the robot captures their attention and attracts them (the \textit{social zone}). As the user and a robot begin a conversation to make a destination and route selection, they are in the \textit{personal zone}. Finally, the hand-off to the wearable device shifts to the \textit{intimate zone}. 

While participants reported that they would prefer not having to use the robot to initiate navigation in their daily life, some participants acknowledged the important role the robot might play as an attractor or social anchor. P3, for example, would like to \textit{``have a humanoid interface to talk with and then keep the results of the interaction''}, similar to P12, who mentioned \textit{``...if I'm at some new place and the robot is there, I would for sure to use it first''}. P8 shared that the change in proximity affected their sense of relationship to the agent differently, as the wearable \textit{``gives the feeling of ownership over something''} while \textit{``feeling like you are asking a favour''} when interacting with the robot.

Still, the participants overwhelmingly affirmed that the robot's role is crucial in special scenarios when they navigate unknown environments, and that navigation with a device in the intimate zone is what they would prefer on a daily basis, potentially enabling cognitive augmentation.

\subsection{\new{Cognitive Augmentation in HCI}}

\new{HCI has a long history of using technology to augment human cognitive capabilities. Theories of cognitive augmentation \cite{clinch2023augmented} describe \textit{directing existing function} as one example of cognitive offloading. Both Schmidt and Clinch \cite{schmidt2017augmenting, clinch2023augmented} highlight how mobile computing, and specifically the use of wearables, can make cognitive enhancements widely accessible. Our study focused on navigation tasks as they are inherently based on our perceptive capabilities \cite{caduff2008assessment}. As such, we hypothesised that reducing the cognitive load of study participants' interactions with a more flexible navigation method, specifically the extension the navigation directions from the stationary interaction to the movement stage would lead to improvements in navigation performance. 

In our case, the navigation tools were conversational tools, but use of traditional (paper or digital) maps, augmented reality devices, or other technologies, might yield similar outcomes. Generative AI, in its different forms, has been shown to be capable of helping humans in everyday tasks \cite{morita2025genaireading, ravishan2024voice, tankelevitch2025tools}. As output reliability and response times improve, it is natural that LLMs, for example, are integrated into more time sensitive and higher stake systems compared to earlier LLMs that have been too inconsistent to find use-cases beyond language processing tasks or entertainment. An interaction such as the human-AI co-navigation might also benefit from the the AI understanding nuanced contextual cues, such as confusion in the user's voice, their movement speed and direction, the crowdedness of the area, the user's apparent language skills and so on. Cognitive Augmentation often benefits from incorporating these contextual cues \cite{xu2025intelligent}. Modern AI systems already make it possible to collect and analyse such data in real time, and it is only a matter of time before wearable devices can run these AI models locally. Our findings show that even simple multi-embodied systems already influence how people structure navigation tasks, suggesting that more capable, context-sensitive AI could amplify these benefits as the underlying technology matures.
}

\subsection{Design Considerations}

Based on our results, we highlight design considerations (RQ3) for a social robot to wearable conversation hand-off, an important step towards bridging the gap in design guidelines for re-embodiment systems \cite{mind_body_identity_2024}, which are promising solutions for assisting humans in various ways~\cite{Luria_Reig_Tan_Steinfeld_Forlizzi_Zimmerman_2019}.

\textbf{Carefully consider the hand-off interaction} as something to be designed with the overall context in mind, and not always simply streamlined technologically.
Our initial thought was that the hand-off simply had to be seamless, to reduce the cognitive and technical load involved in the interaction. 
However, our findings also revealed that some participants experienced the hand-off as clearly fun and in some cases even theatrical ("teleporting assistant"), and that they felt more connected to the AI after a transition.
This could help people rely more on the AI, should that be the design goal in the application in question.
Therefore, paying attention to the transition as something to be designed within the context (marketing, plain throughput of the system, something else) and the intended end-users makes sense.
In our study, we also supported natural language such as "continue on my watch", but the system could certainly be more intelligent and e.g. rely on location-based cues to suggest in different ways a transition as the user walks further away from the robot host, i.e. flexbible triggers.

Thus, \textbf{offering flexible triggers for transition} will cater for diverse user wishes on how to trigger moving the AI agent.
The participants voiced out ideas like hotword detection, and overall, developing different explicit and implicit methods to ensure people know how to trigger the hand-off will be important. This result aligns with \citeauthor{Luria_Reig_Tan_Steinfeld_Forlizzi_Zimmerman_2019}'s findings that user preferences are extremely important in re-embodiment systems \cite{Luria_Reig_Tan_Steinfeld_Forlizzi_Zimmerman_2019}.

Most people preferred to use only the wearable. Thus, we think a natural compromise to strike would be to use the robot to draw awareness that such a navigation possibility even exists, as is often done with similar robots \cite{DeGauquier2018HumanoidShop, Blair2023DevelopmentRobot}, and then perform a \textbf{rapid hand-off as soon as possible}.
Social robots, or indeed any tangible technologies, are excellent due to their attention-drawing characteristics: people have serendipitous encounters with them. 
In other words, robots are seen as engaging but perhaps not quite as functional as a watch that was seen as useful but not inspiring: we may first want to invite the user using the robot to gain attention and then encourage a fast hand-off to a private mobile device.

\subsection{Future Work}

Future work should build on our minimalistic approach to the hand-off interaction and explore its viability beyond indoor navigation tasks. 
For instance, alternative hand-off triggers such as proximity should be explored and different levels of proactivity regarding the hand-off compared.
Further, our experiment included only one robot and one watch, both serving a single purpose. 
Future work should consider multiple robots that serve different locations or even different purposes, potentially even connecting remote locations and even crossing contextual boundaries \cite{Luria_Reig_Tan_Steinfeld_Forlizzi_Zimmerman_2019} rather than a CA serving only locally.
\new{Finally, as \cite{Yamano_Hamajo_Takahashi_Higuchi_2012} suggested, a single-modal interface may be beneficial for navigation tasks. Without hindering flexibility, could be realised with voice commands and proactive, continuous guidance rather than reactive question-answering.}


\subsection{Limitations}

The navigation tasks were small-scale and mostly in one space, which has limited external validity. This lack of variety may have prevented us from encountering specific system benefits or shortfalls. 

Further, the minimalistic system design may have impacted participants' experience of the system as it was missing recommended features \citeauthor{Zahabi_2023} such as interruption the assistant's speech, or showing landmarks visually on the screen in addition to text and speech.  Our system relied on verbal inputs to determine the user's location. A real-world implementation that can use technology-based positioning can help avoid navigation errors and automatically speak the directions. The implementation of the hand-off, while perceived as seamless and reliable, was only experienced in a single task by each participant, and they were not able to gain experience with this unusual way of interaction. 

Finally, as our participant pool was limited to 24 members of the university community, the results may not generalize to broader populations. Our sample consisted primarily of younger, tech-literate individuals, which may have skewed the perceived ease of use and acceptance of the wearable and hand-off conditions. Future work could include individuals with varying technical expertise, older adults, or people with mobility impairments, who may have different needs and preferences for navigational aids.

\section{Conclusion}

We identified indoor navigation as the application area for a re-embodiment system and compared it to two other systems in a user experiment. The wearable-only system was favoured, but participants saw the value in combining different kinds of embodiments, and they were somewhat open to the introduced conversation hand-off technique. This study provides insights for designing multi-embodied public AI assistants and improves our understanding of conversation hand-off challenges and opportunities.

\begin{acks}
This research is partly funded by the Strategic Research Council (SRC), established within the Research Council of Finland (Grants 356128, 335625, 335729), and Research Fellowship funding by the Research Council of Finland (Grants 356128, 349637 and 353790). 
\end{acks}

\bibliographystyle{ACM-Reference-Format}
\bibliography{main}

\appendix


\section{Demographic Questionnaire}\label{sec:appendix_a}

\begin{enumerate}
    \item Age (Text field)
    \item Gender (Female, Male or Other with text field)
    \item Nationality (e.g. Finnish, Hungarian) (Text field)
    \item What is your occupation? If you are a student,  just write student. (Text field)
    \item What do you study? Or if you are a researcher, what is your research topic or department? (Text field)
    \item How often do you rely on some kind of help (map, guidance, signs) for indoors navigation? (5-point scale from "Never" to "Every time")
    \item How familiar are you with social robots, such as Pepper, in general? (5-point scale from "Not at all" to "Extremely familiar")
    \item How familiar are you with AI personal assistants (e.g. Siri, Alexa)? (5-point scale from "Not at all" to "Extremely familiar")
    \item How familiar are you with indoor navigation software? (5-point scale from "Not at all" to "Extremely familiar")
    \item How often do you use indoors navigation software? (Choice between "Less than once a year", "At least once a year", "At least once a month", "At least once a week", "At least once a day")
    \item How often do you use social robots, such as Pepper? (Choice between "Less than once a year", "At least once a year", "At least once a month", "At least once a week", "At least once a day")
    \item How often do you use AI personal assistants (e.g. Siri, Alexa or other) or public AI assistants (e.g. website chatbots, service robots)? (Choice between "Less than once a year", "At least once a year", "At least once a month", "At least once a week", "At least once a day")
\end{enumerate}

\section{Per-condition Questionnaire}\label{sec:appendix_b}

\begin{enumerate}
    \item What devices did you just use? (Choice between "Robot only", "Wearable only", "Both robot and wearable")

    \textbf{NASA Raw (unweighted) Task Load Index (NASA-RTLX) questionnaire:}  
    The NASA Task Load Index is a tool for measuring subjective mental workload. It assesses workload across six dimensions while participants perform a task.

    \item How mentally demanding was the task? (10-point scale from "Not at all" to "Extremely demanding")
    \item How physically demanding was the task? (10-point scale from "Not at all" to "Extremely demanding")
    \item How hurried or rushed was the pace of the task? (10-point scale from "Not at all" to "Extremely")
    \item How successful were you in accomplishing what you were asked to do? (10-point scale from "Not at all" to "Extremely successful")
    \item How hard did you have to work to accomplish your level of performance? (10-point scale from "Not at all" to "Extremely hard")
    \item How insecure, discouraged, irritated, stressed and annoyed were you? (10-point scale from "Not at all" to "Extremely")

    \textbf{Trust Between People and Automation Scale \cite{Jian_Bisantz_Drury_2000}:}  
    How much do you agree with the following statements regarding the system you just tried?

    \item Q1: The system is deceptive (5-point scale from "Not at all" to "Extremely")
    \item Q2: The system behaves in an underhanded manner (5-point scale from "Not at all" to "Extremely")
    \item Q3: I am suspicious of the system's intent, action or outputs (5-point scale from "Not at all" to "Extremely")
    \item Q4: I am wary of the system (5-point scale from "Not at all" to "Extremely")
    \item Q5: The system's actions will have a harmful or injurious outcome (5-point scale from "Not at all" to "Extremely")
    \item Q6: I am confident in the system (5-point scale from "Not at all" to "Extremely")
    \item Q7: The system provides security (5-point scale from "Not at all" to "Extremely")
    \item Q8: The system has integrity (5-point scale from "Not at all" to "Extremely")
    \item Q9: The system is dependable (5-point scale from "Not at all" to "Extremely")
    \item Q10: The system is reliable (5-point scale from "Not at all" to "Extremely")
    \item Q11: I can trust the system (5-point scale from "Not at all" to "Extremely")
    \item Q12: I am familiar with the system (5-point scale from "Not at all" to "Extremely")

    \textbf{Open-ended questions:}
    
    \item What about the system you just tried made it difficult to use for navigation? Name at least two features or characteristics. (Text field)
    \item What about the system you just tried made it easy to use for navigation? Name at least two features or characteristics. (Text field)
    \item Do you want to share anything specific about your experience with the navigation system you just tried? Please do so here. (Text field)
\end{enumerate}

\section{Post-experiment Questionnaire}\label{sec:appendix_c}

\begin{enumerate}
    \item Which case was your preferred navigation method? (Choice between "Instructions from robot only", "Instructions from watch only", "Instructions first from robot, then from watch")
    \item Elaborate on your previous answer (Text field)
    \item How would you suggest that your favourite navigation method is improved?
    \item If "Instructions first from robot, then from watch" was not your favourite method, how do you think it related to the other methods?
    \item If "Instructions first from robot, then from watch" was not your favourite method, how do you think it should be improved?
    \item How do you feel about the AI assistant moving from the robot to your own wearable device?
    \item Do you have any other feedback regarding the experiment or any of the navigation methods? Please share it here.
\end{enumerate}




\section{Final Code Set for Per-Condition Data}\label{sec:appendix_e}

\begin{table}[htbp!]
\centering
\footnotesize
\begin{tabular}{lll}
\toprule
\textbf{positive}        & \textbf{negative}          & \textbf{other}            \\
\midrule
complete\_directions\_good & ai\_made\_mistake          & breakdown\_to\_steps         \\
conv\_transfer\_good       & conv\_transfer\_bad        & cannot\_remember\_steps      \\
direction\_clarity\_good   & hard\_to\_use              & hotword\_wanted              \\
easy\_to\_use              & direction\_clarity\_bad    & landmark\_visuals\_wanted    \\
fast\_directions           & relative\_directions\_bad  & learning\_curve              \\
fun                        & response\_time\_bad        & robot\_gestures\_wanted      \\
mobile\_device\_good       & robot\_gestures\_too\_much & watch\_chat\_history\_wanted \\
privacy\_good              & speech\_rec\_issues        & watch\_gps\_location\_wanted \\
reliability\_good          & speech\_too\_fast          &                              \\
robot\_friendly            & step\_by\_step\_bad        &                              \\
speech\_rec\_good          & task\_understanding\_issue &                              \\
text\_plus\_speech\_good   & technical\_difficulty      &                              \\
text\_plus\_speech\_good   & too\_much\_reading         &                              \\
\textbf{}                  & too\_time\_consuming       &                              \\ \bottomrule
\end{tabular}
\end{table}

\newpage

\section{Final Code Set for Post-Experiment Data}\label{sec:appendix_f}

\begin{table}[htbp!]
\centering
\footnotesize
\begin{tabular}{ll}
\toprule
\textbf{positive}                      & \textbf{negative}               \\ \midrule
watch\_better\_confidence              & bending\_down\_bad              \\
ai\_on\_wearable\_interesting          & conv\_handoff\_pointless        \\
direction\_clarity\_good               & conv\_transfer\_bad             \\
easy\_to\_use                          & human\_element                  \\
fun\_to\_use                           & relative\_directions\_bad       \\
mistakes\_are\_fine                    & robot\_pointless                \\
mobile\_device\_good                   & static\_robot\_bad              \\
reliability\_good                      & visual\_map\_wanted             \\
robot\_ease\_anxiety                   &                                 \\
speech\_rec\_good                      &                                 \\
step\_by\_step\_good                   &                                 \\
watch\_sense\_of\_ownership            &                                 \\ \toprule
\textbf{request}                       & \textbf{other}                  \\ \midrule
better\_consistency\_wanted            & new\_vs\_returning\_visitor     \\
gps\_location\_wanted                  & one\_device\_better\_than\_two  \\
hotword\_wanted                        & test\_round\_wanted             \\
initiate\_handoff\_from\_watch\_wanted & would\_download\_as\_phone\_app \\
list\_output\_wanted                   & handoff\_better\_than\_robot    \\
make\_handoff\_seamless                &                                 \\
make\_ui\_fancier                      &                                 \\
more\_language\_flexibility\_wanted    &                                 \\
watch\_chat\_history\_wanted           &                                 \\ \bottomrule
\end{tabular}
\end{table}

\clearpage

\section{Layout of the experiment area}\label{sec:appendix_f}

\begin{figure}[htpb!]
    \centering
    \includegraphics[width=0.75\textwidth]{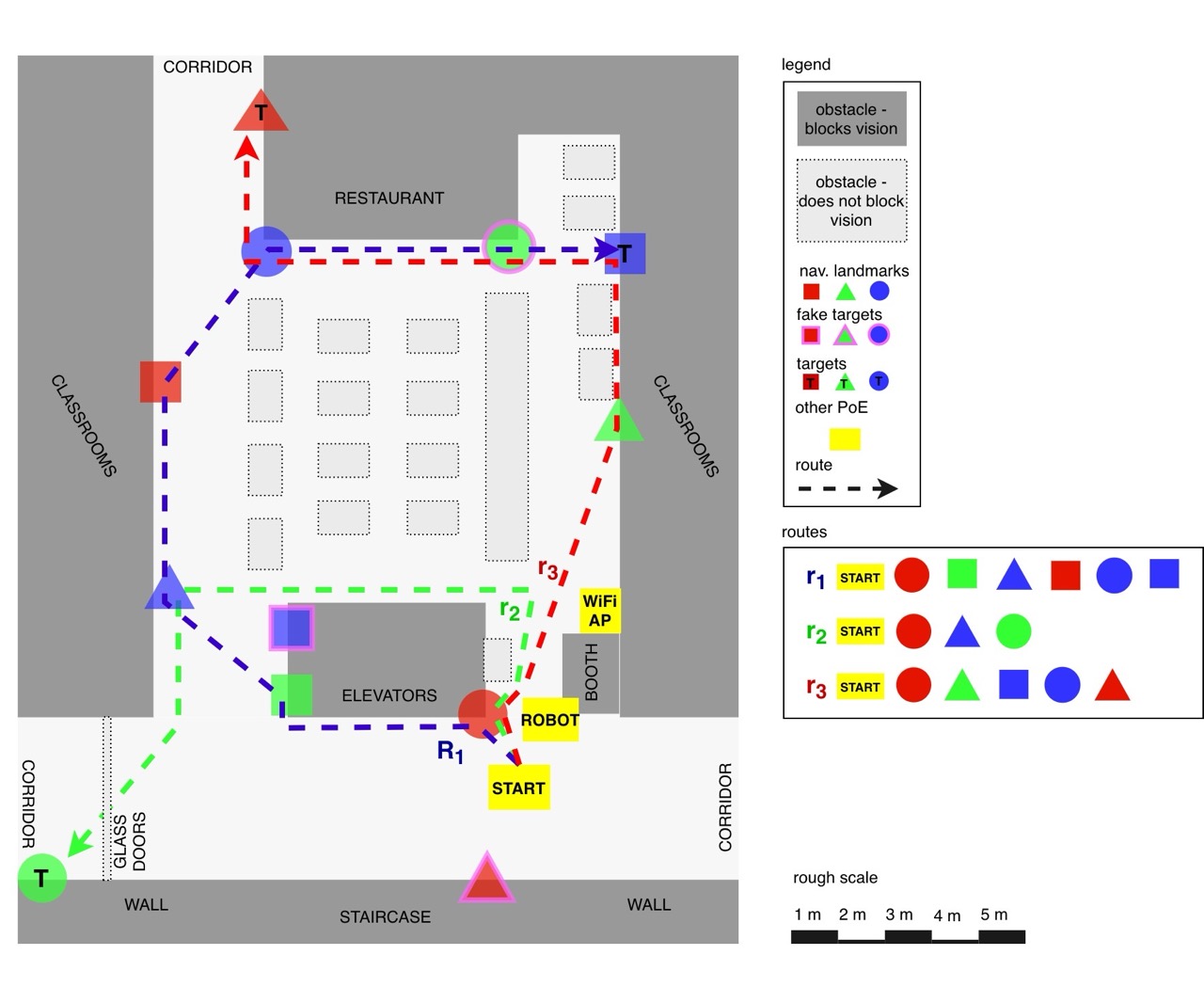}
    \caption{Layout of the experiment area, navigation routes and landmarks.}
    \label{fig:routes}
\end{figure}




\end{document}